%% file: MuonDump.tex
\pdfoutput=1
\documentclass[preprint,3p]{elsarticle}

\usepackage{amssymb}
 \usepackage{lineno}
 \usepackage{color}

\journal{Nuclear Instruments and Methods in Physics Research Section A}

\begin{document}

\begin{frontmatter}

%% Title, authors and addresses

%% use the tnoteref command within \title for footnotes;
%% use the tnotetext command for theassociated footnote;
%% use the fnref command within \author or \address for footnotes;
%% use the fntext command for theassociated footnote;
%% use the corref command within \author for corresponding author footnotes;
%% use the cortext command for theassociated footnote;
%% use the ead command for the email address,
%% and the form \ead[url] for the home page:
%% \title{Title\tnoteref{label1}}
%% \tnotetext[label1]{}
%% \author{Name\corref{cor1}\fnref{label2}}
%% \ead{email address}
%% \ead[url]{home page}
%% \fntext[label2]{}
%% \cortext[cor1]{}
%% \affiliation{organization={},
%%             addressline={},
%%             city={},
%%             postcode={},
%%             state={},
%%             country={}}
%% \fntext[label3]{}

\title{Measurement of muon flux behind the beam dump \\ of the J-PARC Hadron Experimental Facility}

%% use optional labels to link authors explicitly to addresses:
%% \author[label1,label2]{}
%% \affiliation[label1]{organization={},
%%             addressline={},
%%             city={},
%%             postcode={},
%%             state={},
%%             country={}}
%%
%% \affiliation[label2]{organization={},
%%             addressline={},
%%             city={},
%%             postcode={},
%%             state={},
%%             country={}}

\author[NDA]{T.~Matsumura\corref{ctr1}}
\ead{toru@nda.ac.jp}
\cortext[ctr1]{Corresponding author. Tel.: +81-46-841-3810; fax: +81-46-844-5912.}
\author[NDA]{Y.~Hirayama}
\author[KEK]{G.Y.~Lim}
\author[Osaka]{H.~Nanjo}
\author[KEK]{T.~Nomura}
\author[KEK]{K.~Shiomi}
\author[KEK]{H.~Watanabe}

\affiliation[NDA]{organization={Department of Applied Physics, National Defense Academy of Japan},%Department and Organization
%            addressline={1-10-20 Hashirimizu}, 
            city={Yokosuka},
            postcode={239-8686}, 
            state={Kanagawa},
            country={Japan}}
\affiliation[KEK]{organization={Institute of Particle and Nuclear Studies, High Energy Accelerator Research Organization (KEK)},%Department and Organization
            city={Tsukuba},
            postcode={305-0801}, 
            state={Ibaraki},
            country={Japan}}
\affiliation[Osaka]{organization={Department of Physics, Osaka University},%Department and Organization
            city={Toyonaka},
            postcode={560-0043}, 
            state={Osaka},
            country={Japan}}

\begin{abstract}
%% Text of abstract
A muon-flux measurement behind the beam dump 
of the J-PARC Hadron Experimental Facility was performed with 
a compact muon detector 
that can be inserted into a vertical observing hole with 81~mm in diameter
which was dug underground.  
The detector consists of 12 plastic scintillation strips with a length of 80~mm each, 
which are arranged with cylindrical shape and contained inside an aluminum housing with an outer diameter of 75~mm. 
A silicon photomultiplier is coupled to the end of each strip to collect the scintillating light.
The flux of the muons penetrating the beam dump was scanned vertically at intervals of 0.5~m, 
showing a wide distribution with a maximum at the beam level.  
The muon flux was consistent with the expectation from a Monte-Carlo simulation 
at more than 1~m away from the beam axis,  
which is expected to be used for signal-loss evaluation in the future KOTO~II experiment 
for measuring rare kaon decays.
The data can also be used in improving 
the accuracy of shielding calculations in the radiation protection.
\end{abstract}

%%Graphical abstract
%\begin{graphicalabstract}
%\includegraphics{grabs}
%\end{graphicalabstract}

%%Research highlights
%\begin{highlights}
%\item Research highlight 1
%\item Research highlight 2
%\end{highlights}

\begin{keyword}
%% keywords here, in the form: keyword \sep keyword
muon flux \sep beam dump \sep proton accelerator \sep J-PARC \sep KOTO

%% PACS codes here, in the form: \PACS code \sep code

%% MSC codes here, in the form: \MSC code \sep code
%% or \MSC[2008] code \sep code (2000 is the default)

\end{keyword}

\end{frontmatter}

%\linenumbers

%% main text
\input{Section1.tex}

\input{Section2.tex}

\input{Section3.tex}

\input{Section4.tex}

\input{Section5.tex}

\input{Section6.tex}

\section{Summary}
We performed a muon-flux measurement in the observing hole 
just behind the beam dump of the J-PARC Hadron Experimental Facility.
Using a compact muon detector, the vertical muon-flux distribution  
with an interval of 0.5~m was measured. 
A calculation with GEANT4 simulation was almost consistent with 
experimental data at positions more than 1~m away from the beam axis.  
The results enable a more reliable evaluation of the signal loss 
in the future KOTO-II experiment.

\section*{Acknowledgement}
We gratefully acknowledge the support of the staff at J-PARC
for providing excellent experimental conditions. Part of this work 
was financially supported by JSPS KAKENHI Grant Numbers JP21H04483, JP21H04995.

%% The Appendices part is started with the command \appendix;
%% appendix sections are then done as normal sections
%% \appendix

%% \section{}
%% \label{}

%% For citations use: 
%%       \citet{<label>} ==> Jones et al. [21]
%%       \citep{<label>} ==> [21]
%%

%% If you have bibdatabase file and want bibtex to generate the
%% bibitems, please use
%%
%\bibliographystyle{elsarticle-num-names} 
%\bibliography{MyBibtex}

%% else use the following coding to input the bibitems directly in the
%% TeX file.

\end{document}

%% file: Section1.tex
\section{Introduction}

This article reports a muon-flux measurement behind the beam dump of an experimental facility at a 30-GeV proton accelerator. 
The measurement was motivated by a future kaon rare-decay experiment at Japan Proton Accelerator Research Complex (J-PARC)~\cite{nagamiya2012}. 
In this section, we describe the research background of this study.

The KOTO experiment is currently being conducted at the J-PARC Hadron Experimental Facility (HEF) 
to search for the $K_{L}\rightarrow\pi^0\nu\bar{\nu}$ decay~\cite{comfort2006,yamanaka2012}. 
The branching ratio of $K_{L}\rightarrow\pi^0\nu\bar{\nu}$ is predicted to be  
$(2.94\pm0.15)\times 10^{-11}$ and is extremely rare in the Standard Model (SM)~\cite{buras2023}. 
The accurate theoretical prediction of this process allows us to probe new physics beyond the SM.
The current upper limit of the branching ratio was set 
by KOTO to be $3.0\times 10^{-9}$~\cite{ahn2019}. 
KOTO plans to continue the search toward a sensitivity better than $10^{-10}$, 
however will not reach the sensitivity that can observe the signal events at the sensitivity predicted by SM.  
To discover the $K_{L}\rightarrow\pi^0\nu\bar{\nu}$ decay and measure its branching ratio, 
we are planning the new experiment KOTO~II, which can observe $\sim$35 SM signal events~\cite{nanjo2023,aoki2021}, 
as a part of the extension project of HEF~\cite{aoki2021}.

In both the KOTO and KOTO~II experiments, 
we employ a calorimeter to detect $\pi^0\to \gamma\gamma$ and 
a large volume of veto %VETO 
counters covering the entire decay region to ensure no other particles in the final state. 
One concern for a measurement with veto %VETO 
counters is a signal loss  
due to an accidental hit in a veto %VETO 
counter.  
An estimation with a current design of the KOTO~II experiment shows that the signal loss  
reaches 39\% due to accidental hits caused by 
$K_L$ decays and beam particles including photons and neutrons~\cite{aoki2021}.
Thus, it is crucial to suppress signal losses from any other sources as much as possible.

\begin{figure*}[h]
\begin{center}
\includegraphics[width=15cm, bb=0 0 960 540, trim=0cm 5cm 0cm 5cm, clip]{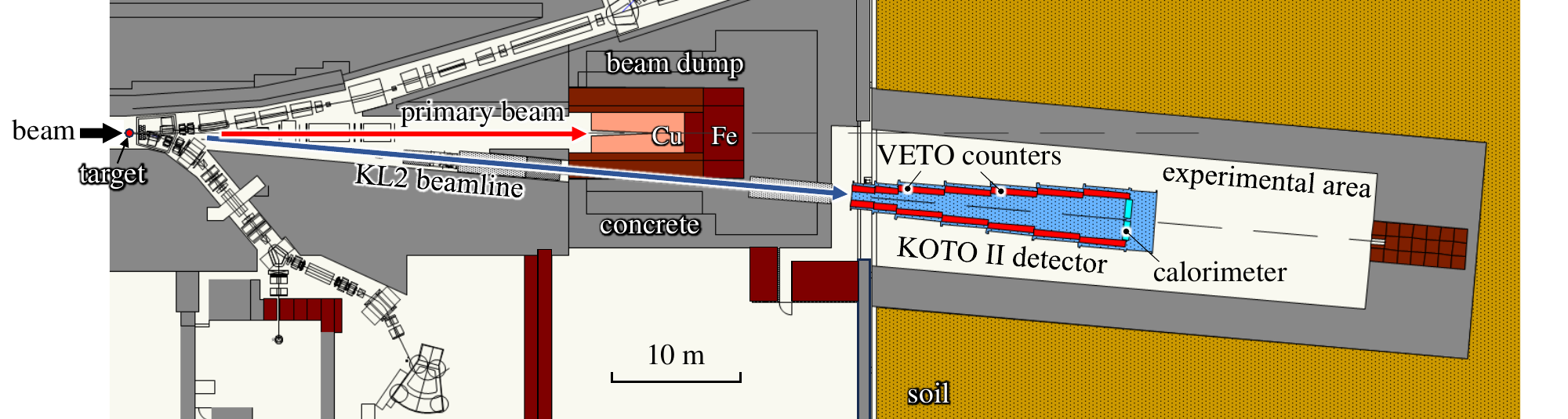}
\caption{Current design of the neutral beamline (KL2) and the experimental area for the KOTO II experiment (plan view).  
The 30-GeV proton beam is injected onto the production target and secondary particles are transported to several beamlines including the KL2 beamline directed to 5 degrees. 
The primary protons passing through the target are absorbed in the beam dump.}
\label{fig:beamline}
\end{center}
\end{figure*}

Figure~\ref{fig:beamline} shows the current design of the neutral beamline (KL2) 
and experimental area for KOTO~II in the HEF extension project. 
The direction of the neutral beamline to that of the primary beam is 5 degrees, 
which is further forward compared to the 16 degrees in the current beamline for the KOTO experiment,  
so as to obtain a larger $K_L$ flux. 
We optimized the beamline length to be 43~m, 
considering both the reduction of short-lived particles such as $\Lambda$ and $K_S$  
coming from the target and the solid angle with a given beam size at the experimental area.  
Hence, the location of the KOTO~II detector would be downstream of the beam dump, 
which is placed at the end of the primary beamline to absorb protons passing through the production target.

The current beam dump of HEF is composed of 
a copper core with a conical hole at its center, and surrounding iron and concrete blocks for radiation shielding. 
The dump will be used also in the extended facility by moving the location further downstream. 
Most particles generated by interactions between the protons and the materials in the beam dump 
are absorbed by the dump itself and shields. 
However, high-energy muons originating from charged-pion decays 
potentially penetrate the radiation shields and can reach the KOTO~II detector 
and cause accidental hits in it. 
In order to evaluate the effects, we performed a 
muon-flux measurement behind the beam dump at the current facility.  
Since the beam height is below the ground level, we dug a hole for observation and 
measured a vertical distribution of the muon flux with a compact detector 
developed for this purpose.

We first describe the experimental method and the muon detector in Section~2. 
%In the next section, the experimental method and the muon detector are described. 
This detector can also measure fast neutrons at the same time; 
however, in this article, we only report on muon measurements and  
the fast-neutron measurements will be discussed elsewhere.
In Section~3, data analysis including energy calibration and counting rate estimation is discussed.  
The systematic uncertainties of the muon flux are evaluated in Section~4. 
The results and comparison with a simulation will be discussed in Section~5 and 6.

%% file: Section2.tex
\section{Experimental method}

\subsection{Measurement location}

\begin{figure}
\begin{center}
\includegraphics[width=9cm, bb=0 0 960 540, trim=3cm 0cm 3cm 0cm, clip]{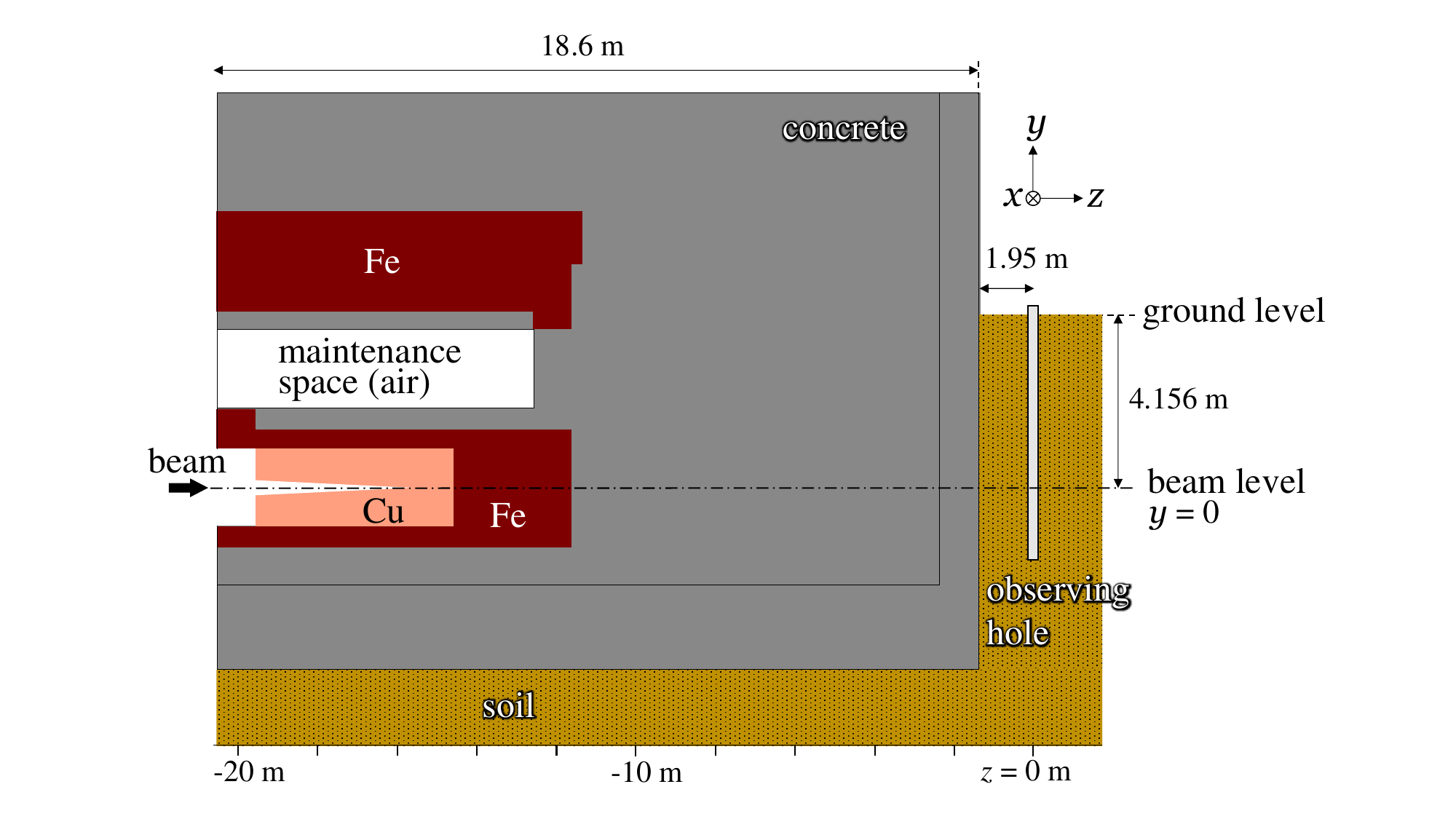}
\caption{Illustration of the side cross-sectional view of the current beam dump of the J-PARC Hadron Experimental Facility 
and the location of the observing hole. The directions of the coordinate system defined in the text are shown. 
The origin of the system is set to be at the position of the observing hole on the primary beam axis.}
\label{fig:meas_loc}
\end{center}
\end{figure}

The flux measurement was performed at an observing hole located 1.95~m downstream along the primary beam axis 
from the downstream end of the beam dump of the current facility, as shown in Fig.~\ref{fig:meas_loc}. 
The level of the primary proton beam was 4.156~m under the ground level.
The depth of the hole was 6.15~m from the ground. 
A 4-mm-thick stainless-steel pipe with an inner diameter of 81~mm was buried in the hole. 
We defined the $x$ and $y$ coordinates as horizontal and vertical positions, respectively, 
whose origin is set to be on the primary beam axis, and the $z$ axis to be the primary beam direction. 
Detailed descriptions of the facility and the beam dump are provided in Ref.~\cite{agari2012}.

\subsection{Beam condition} \label{subsec:beam}

The 30-GeV beam was extracted from the proton synchrotron during 2 seconds 
with a 5.2-second repetition cycle.
The beam intensity delivered to the target during the measurement was 30 kW.   
This corresponds to $1.7\times10^{13}$ protons impinging the beam dump
by assuming 50\% beam loss in the 66-mm-long gold target~\cite{saito2022}, 
which was placed 47.4~m upstream from the beam dump 
(and is not shown in Fig.~\ref{fig:meas_loc}).
The standard deviation of the beam size at the entrance of the beam dump was measured  
to be 3.0~cm using a residual gas ionization profile monitor (RGIPM)~\cite{agari2012}. 
As described in the next section, 
this information was used as input parameters for a simulation 
reproducing secondary particles generated from the beam dump. 
 
\subsection{Expected intensity of secondary particles at the observing hole} \label{subsec:expectation}

Prior to the measurement, we evaluated the secondary particles 
reaching the location by a simulation code using GEANT4 version 10.5.1~\cite{agostinelli2003,allison2006} 
with a hadron-physics package of QGSP\_BERT~\cite{allison2016}.  
In this code, we defined the detailed geometry of the beam dump structure 
including the copper core, iron shield, concrete shield, and 
maintenance space for cooling water flow as shown in Fig.~\ref{fig:meas_loc}.  
This simulation is referred to as the ``beam-dump simulation'' in this article.  

\begin{figure}
\begin{center}
\includegraphics[width=9cm, bb=0 0 960 540, trim=6.5cm 1.8cm 6.5cm 2.cm, clip]{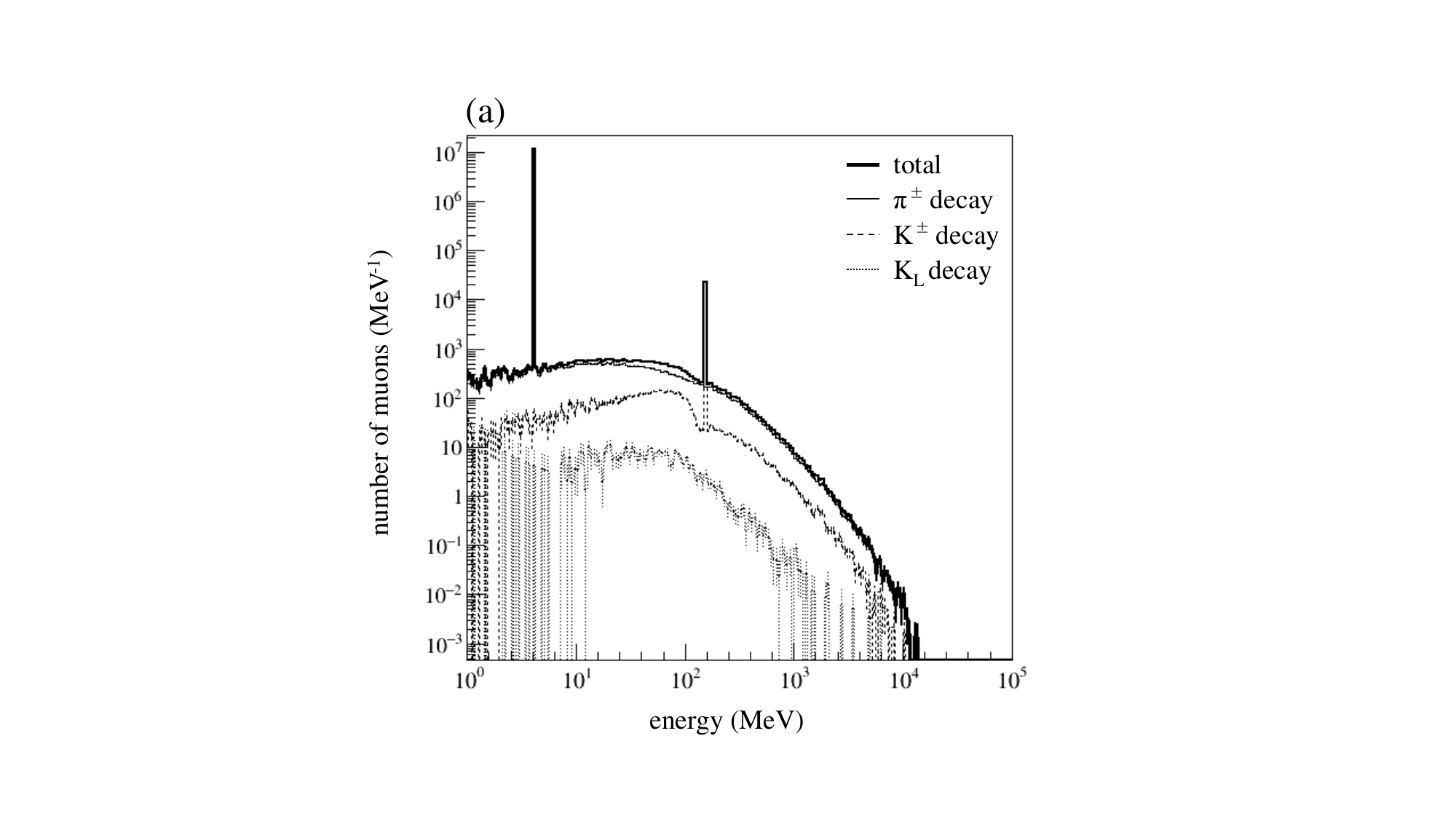}
\includegraphics[width=9cm, bb=0 0 960 540, trim=6.5cm 3.cm 6.5cm 4.5cm, clip]{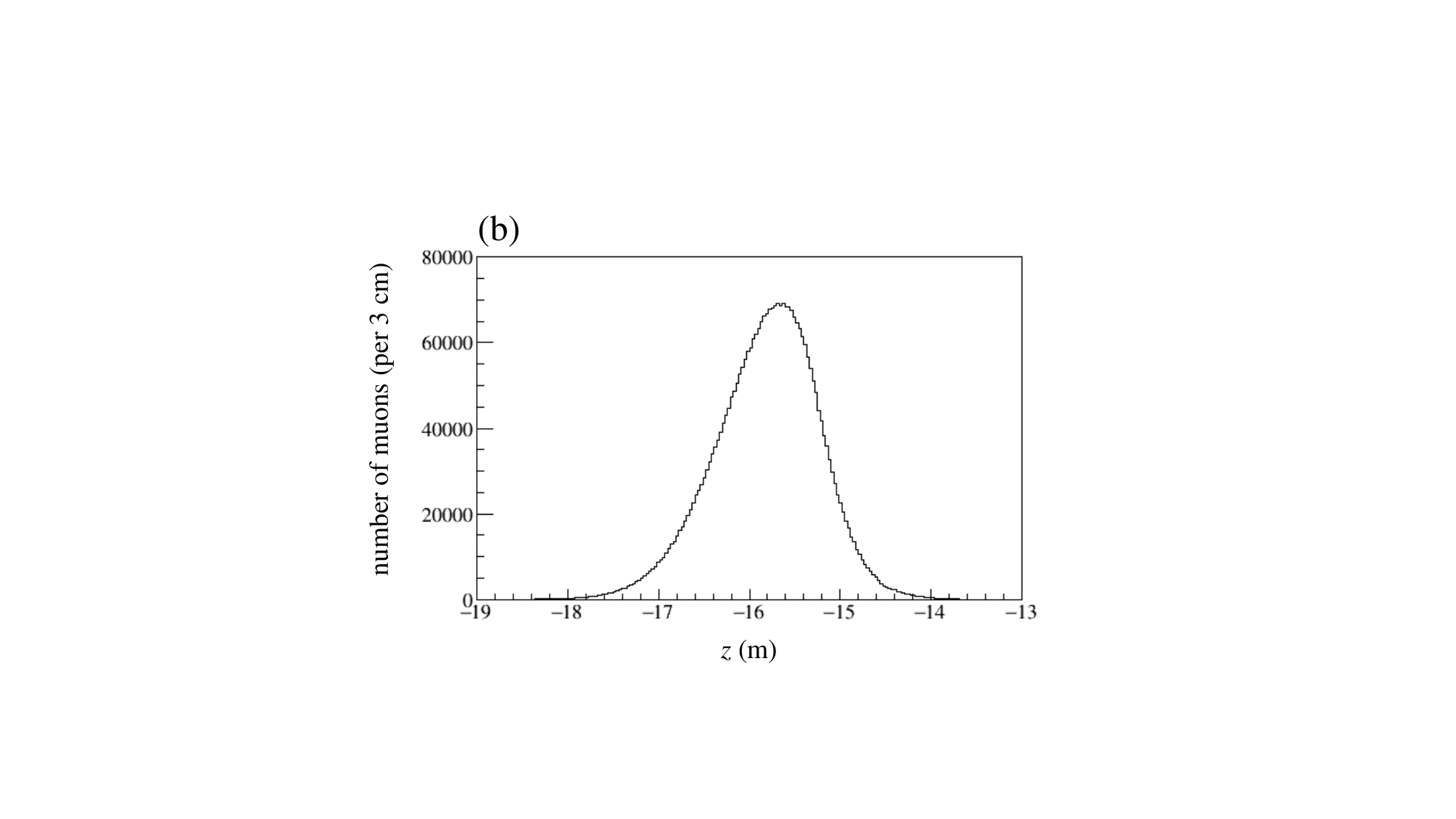}
\caption{Simulation results of all the muons produced in the beam dump in the case of 10$^6$ of 30~GeV proton injection. 
(a) Production energy spectra of the muons 
with the breakdown of the production processes through the three types of hadronic decays.  
Two sharp peaks correspond to $\pi\to\mu\nu$ decay of stopped pions (4.1~MeV) and $K\to\mu\nu$ decay of 
stopped kaons (152~MeV).  
(b) Position distribution of the produced muons in the $z$ direction. 
The relationship between $z$ position and geometry can be seen in Fig~\ref{fig:meas_loc}.}
\label{fig:prod_mu}
\end{center}
\end{figure}

Figure~\ref{fig:prod_mu}(a) shows energy spectra %and $z$ positions 
of all the muons produced in the beam dump. 
Most of the muons originate from $\pi$ and $K$ decays. % from Fig.~\ref{fig:prod_mu}(a). 
On average, 3.2 muons are generated per incident proton in the beam dump.   
Among them, a fraction of $1.3\times10^{-6}$ of the muons reach the location of the observing hole at $z=0$,  
and most of the other muons stop in the shield because the average energy loss in the shielding materials is 11.8~GeV.  
As shown in Fig.~\ref{fig:prod_mu}(b), the $z$ position of the muon production 
is $-15.8$~m on average with a spread of 0.6~m in the root-mean-square.
This information will be used later in the analysis of the detector acceptance.

\begin{figure}
\begin{center}
\includegraphics[width=9cm, bb=0 0 960 540, trim=6.5cm 2.cm 6.5cm 3.cm, clip]{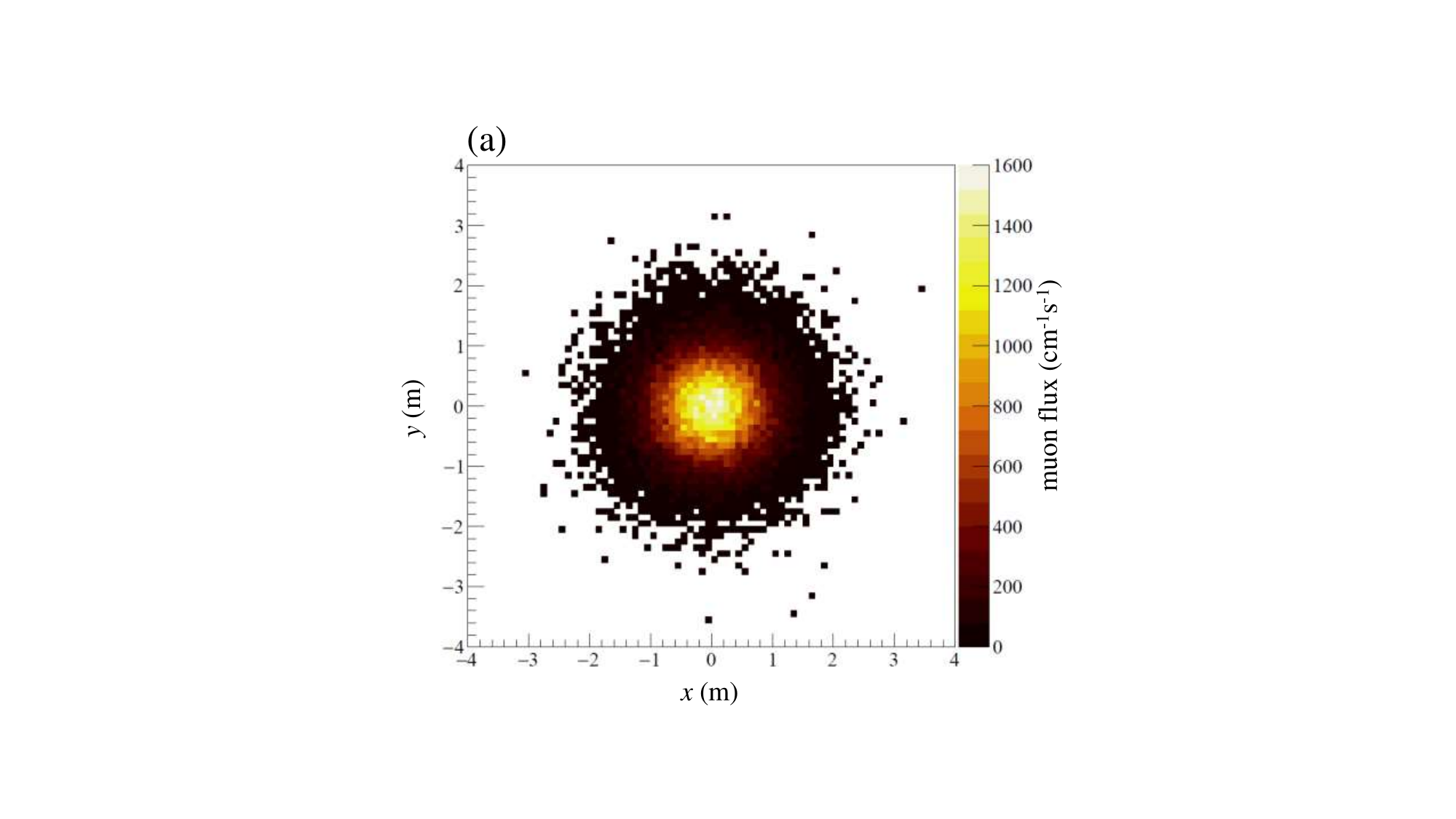}
%\end{center}
%\begin{center}
\includegraphics[width=9cm, bb=0 0 960 540, trim=6.5cm 2.cm 6.5cm 2.8cm, clip]{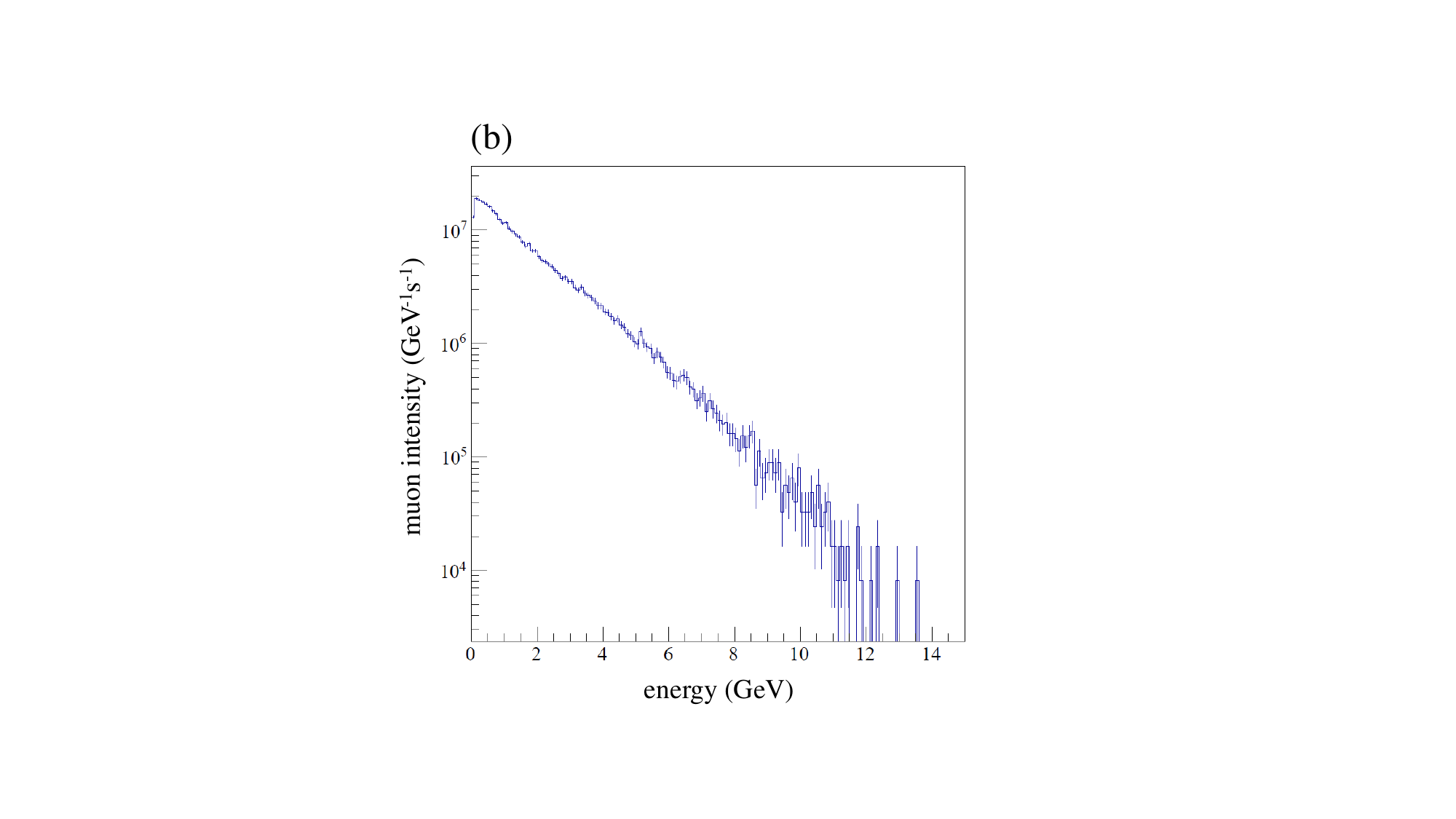}
\caption{Simulation results with the 30~kW beam operation for (a) muon-flux distribution on the plane perpendicular to the beam axis at the observing hole 
and (b) energy distribution of muons passing through the plane.}
\label{fig:dump_sim}
\end{center}
\end{figure}

Figure~\ref{fig:dump_sim}(a) shows the muon-flux distribution on the $x$-$y$ plane at the $z$ position of the observing hole ($z=0$),  
calculated by the beam-dump simulation.  
The distribution has a peak at the center and is symmetrical around the beam axis.
Figure~\ref{fig:dump_sim}(b) shows the energy distribution of the muon, showing an exponential slope with an average energy of 1.7~GeV.
%The average production point of the muons along the beam axis was at $z=-16$~m, 
%which will be used later in the analysis of the detector acceptance. 
The intensity of the muons during the beam extraction integrated over the whole $x$-$y$ plane at $z=0$    
was estimated to be $3.5\times 10^7$~Hz with the 30~kW beam.

In addition to the muons, $\gamma$ and $e^\pm$ and a smaller amount of neutrons were also observed in the simulation.   
$\gamma$'s and $e^\pm$'s were originated from $\delta$-rays associated with the ionization by muons 
and the electromagnetic shower induced by the $\mu^{\pm}\to e^\pm \nu\bar{\nu}$ decay. 
Neutrons were mainly produced by nuclear interactions between negative muons and materials 
near the observing hole, such as soil and the detector itself.  
Namely, these $\gamma$, $e^\pm$, and neutrons did not directly come from the core of the beam dump, 
but were associated with muons that penetrated the beam dump. 
The estimated rates of $\gamma$, $e^\pm$ and neutrons  
were $2.3\times 10^7$~Hz, $4.3\times 10^6$~Hz, and $1.8\times10^5$~Hz, respectively, 
integrated over the whole $x$-$y$ plane at $z=0$.

%\begin{table}
%\caption{position dependence of the particle flux} \label{tab:MC_flux}
%\begin{center}
%\begin{tabular}{c|cccc}
%\hline
%$y$ (m)  & $\mu^{\pm}$ & $n$ & $\gamma$ & $e^{-}$($e^+$) \\ \hline
%0 & $1431\pm 22$  & $6.4\pm2.2$ & $1547\pm 36$ & $132\pm 18$   \\
%1 & $370\pm 11$  &  $4.9\pm1.7$ & $384\pm 18$ & $16\pm 56$   \\
%2 & $13.6\pm 2.1$  &  -  & $13\pm 3$ & $0.8\pm  0.8$   \\ \hline
%\end{tabular}
%\end{center}
%\end{table}
%
%Table~\ref{tab:MC_flux} shows particle flux and average energy expected from the simulation in the vertical position $y$. 
%Based on the simulation, the distribution in the $x$-$y$ plane for particles other than muon is expected to be similar to that of muon.

\subsection{Detector to measure the flux}

\begin{figure}
\begin{center}
\includegraphics[width=9cm, bb=0 0 980 540, trim=8cm 0cm 6cm 0cm, clip]{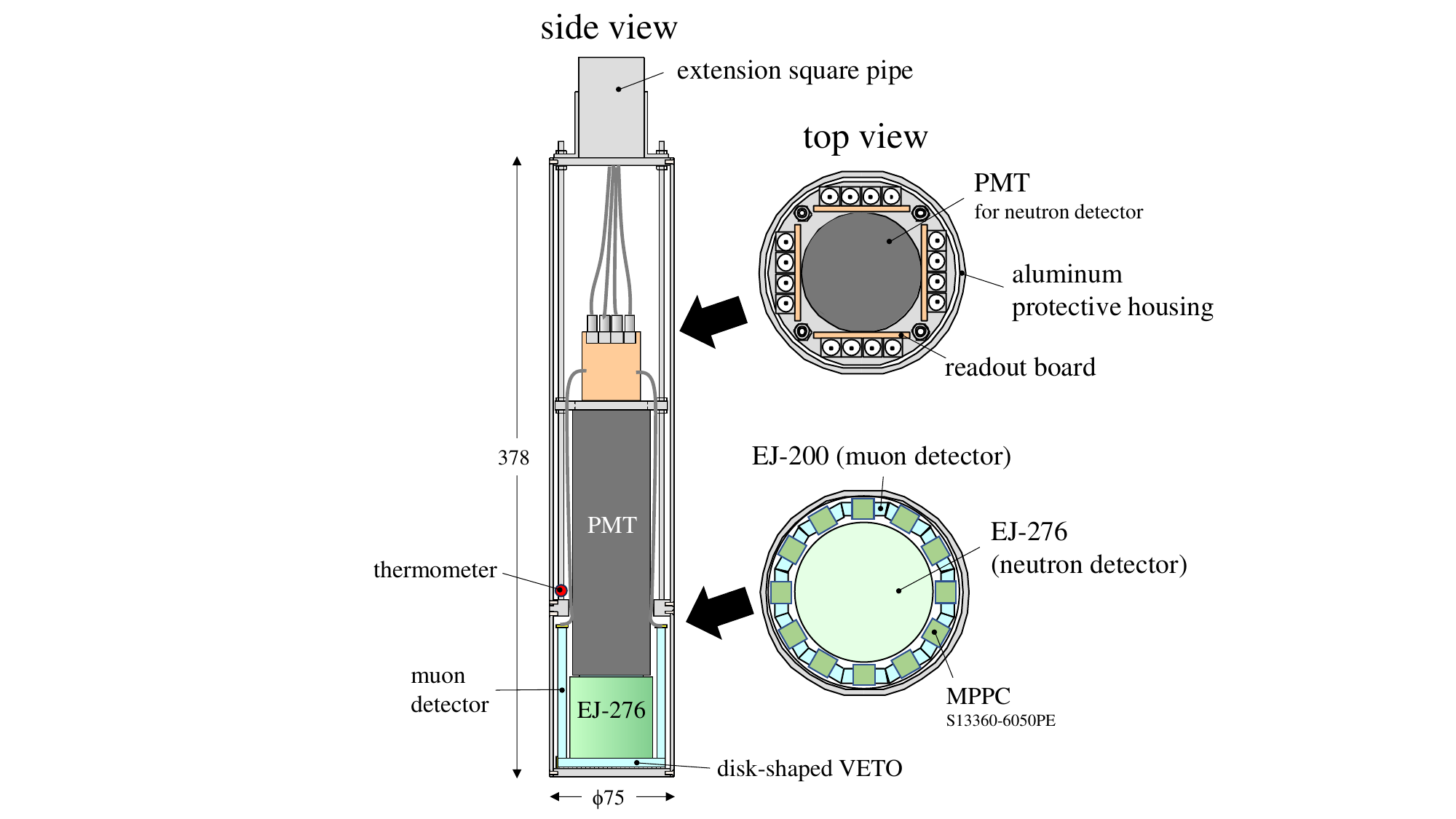}
\caption{Compact detector system for the muon and fast neutron measurements. 
The cylindrical neutron detector (EJ-276) and 12 strips of 
the trapezoidal muon detector (EJ-200) were packaged inside the aluminum protective housing.}
\label{fig:detector}
\end{center}
\end{figure}

We developed a compact detector that can be inserted into the observing hole 
to measure the intensity distribution of muons, as shown in Fig.~\ref{fig:detector}.   

Fast neutrons can be detected using a 2-inch cylindrical plastic scintillator, ELJEN EJ-276, 
suitable for performing n/$\gamma$ separation through the pulse shape discrimination (PSD) technique. 
A Hamamatsu H10828 photomultiplier tube (PMT), which has a 1.5-inch diameter, 
was attached to the scintillator.
We refer to this detector as the ``neutron detector''.
Note that data from the neutron detector is not employed in this article.  

For muon detection, 
12 plastic scintillator strips with trapezoidal cross section, 
ELJEN EJ-200, were placed in a cylindrical shape around the neutron detector.
Each strip was 80~mm long, 14.7~mm wide on the short side, and 5~mm thick.
We covered each strip with two layers of a Teflon reflector sheet, then 
wrapped the reflector with aluminum foil to prevent light leak to the adjoining scintillators. 
A 6-mm-square MPPC, Hamamatsu S13360-6050PE, was directly attached to 
the top end of each strip via optical cement, ELJEN EJ-500.
Twisted pair cables soldered to the MPPC were connected to the readout board placed on top  
and was used for bias voltage supply and signal readout. 
A common bias voltage of $-$54.70~V was applied to all 12 MPPC with a source meter, KEITHLEY model 2400.
The gain variation among the MPPC's was $\pm27$\%,   
and the gain correction was applied in the offline analysis. 
This set of 12 scintillator strips is referred to as the ``muon detector''.
As described later, the muon flux was determined by coincidence measurement of two facing strips. 

A disk-shaped plastic scintillator at the bottom (EJ-200) was used as a charged-particle veto counter, 
together with the muon detector, 
when measuring neutrons. 
The disk was 68~mm in diameter and 5~mm in thickness. 
The same type of MPPC used for the muon detector was attached to its one side for readout.

All detection devices were contained inside an aluminum protective housing 
with an outer diameter of 75~mm and a thickness of 2~mm.
A platinum resistance thermometer was installed to measure the temperature 
in the housing near the MPPC using a digital multimeter, Agilent 34410A. 
The signal from the detector was sent to readout devices
in the counting room via 10-m thin coaxial cables (RG-174) and 30-m RG-58C/U cables.

%\begin{figure}
%\begin{center}
%\includegraphics[width=7cm, bb=0 0 960 540, trim=8cm 0cm 9cm 0cm, clip]{fig5.pdf}
%\caption{Schematic view of the data readout. All the signal cables from the detector were connected to a 500~MHz digitizer. 
%The signal from the neutron detector was split into two channels, and 100 samples of 14-bit waveform data were recorded;  
%one of the two channels was attenuated to 1/16 to acquire data with a high-dynamic range.
%A PC retrieved the stored data through a LAN cable.}
%\label{fig:readout}
%\end{center}
%\end{figure}

%As shown in Fig.~\ref{fig:readout}, 
Signals from the detector system were digitized 
with a 16-channel 500~MHz digitizer (TechnoAP,  APV8516-14).
A desktop computer retrieved the stored data through a LAN cable.
The recorded data have time stamps with a resolution of 7.8 ps that can be used for offline analysis to determine coincidence.
The online energy-threshold of the muon detector was from 0.03 to 0.10~MeV depending on  the gain of each strip. 
The digitizer additionally received the timing signal from the accelerator every 5.2 seconds, 
synchronizing with the beam extraction.
This enables us to determine whether the recorded signal was made during the beam extraction period of 2 seconds.

\subsection{Detector response}
\label{subsec:response}

\begin{figure}
\begin{center}
\includegraphics[width=9.0cm, bb=0 0 960 540, trim=3.5cm 1.0cm 2cm 1.5cm, clip]{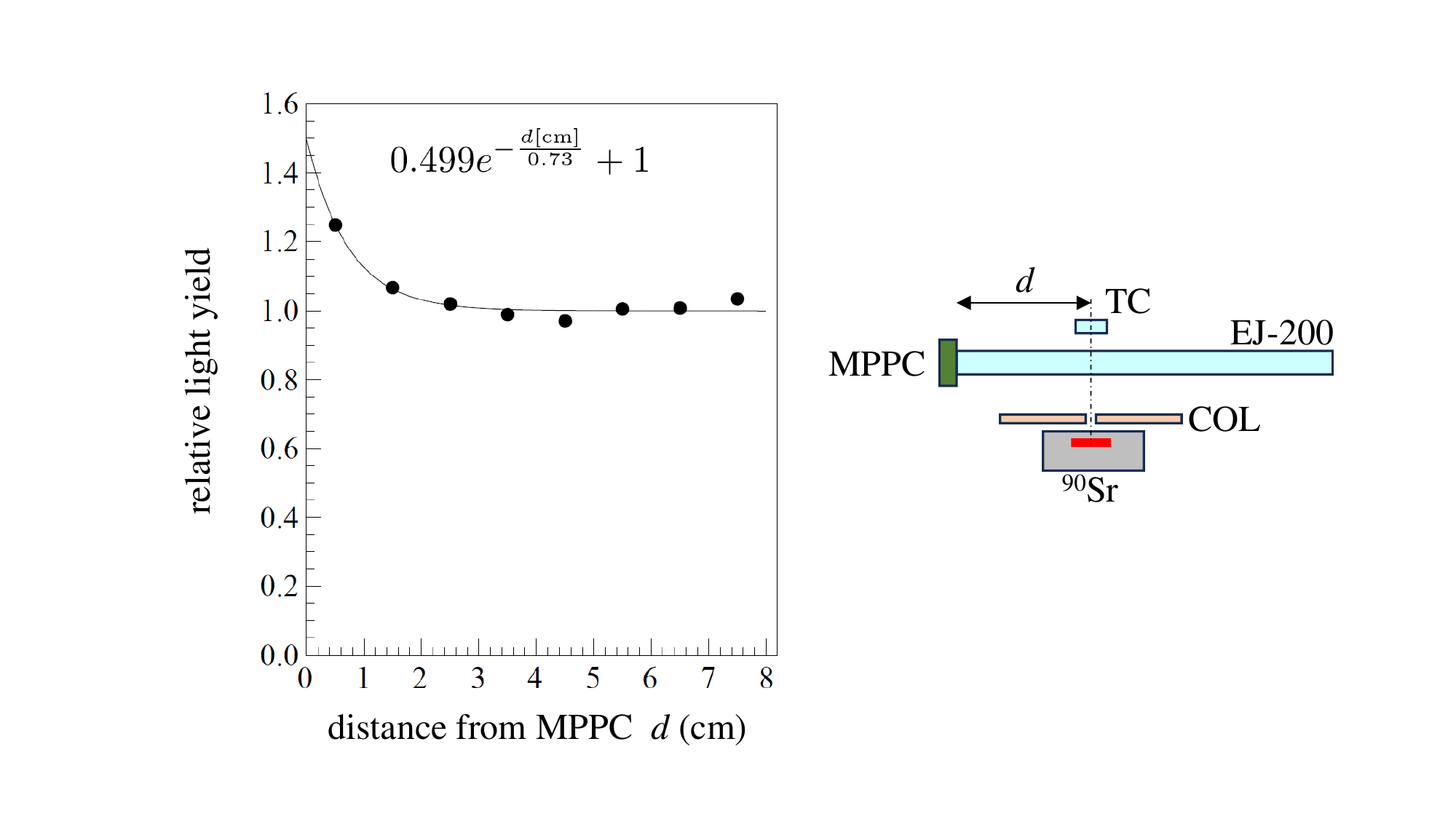}
\caption{Position dependence of the light yield of an 8-cm-long muon detector strip. 
An illustration at the right of the graph shows the setup: $1\times 1{\rm\ cm}^2$ trigger counter (TC), 
scintillator strip (EJ-200), a 2-mm-thick copper collimator (COL, $\phi$5~mm), and a $^{90}$Sr $\beta$-ray source. 
The exponential function fitted to the data was used as an input of the detector simulation. }
\label{fig:pos_dep}
\includegraphics[width=9.5cm, bb=0 0 960 540, trim=3cm 1.0cm 3cm 1.5cm, clip]{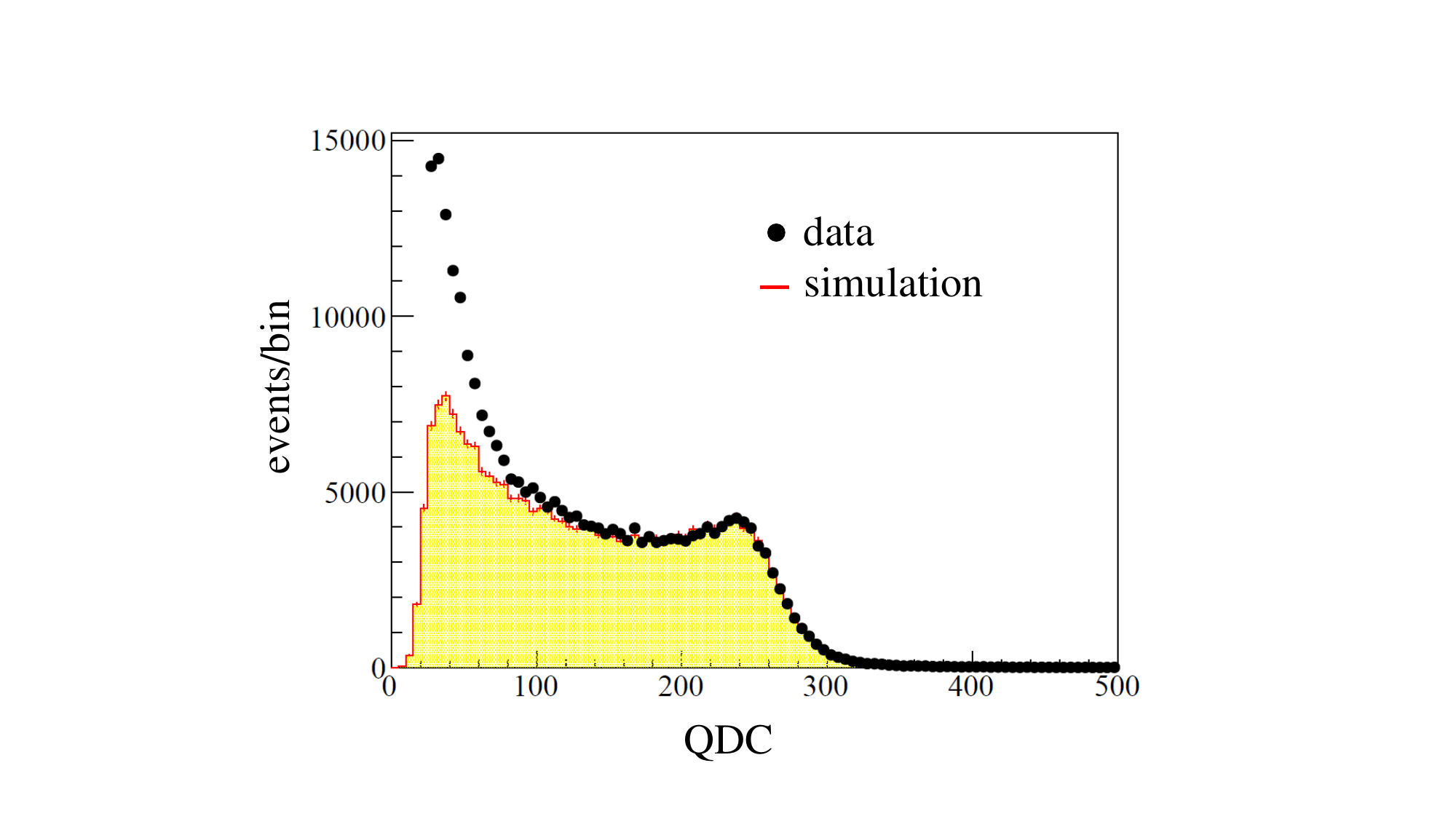}
\caption{Detector response of the muon detector in the $\gamma$-ray measurement from $^{137}$Cs. 
The measurement was performed on the ground. 
The low energy part of the data is dominated by scattered $\gamma$-rays, which is not considered in the detector simulation.}
\label{fig:Cs137}
\end{center}
\end{figure}

The position dependence of the light yield of the muon detector was 
evaluated, prior to the flux measurement, by using a collimated $\beta$ source, 
as shown in Fig.~\ref{fig:pos_dep}.  
The light collectivity became slightly higher in the region closer to the MPPC.  
The fitted function in the figure was used as a response function in the detector simulation.

A detector simulation code was prepared to reproduce the detector response with the GEANT4 toolkit,  
in which all detector components were defined except for the vacuum tube of the PMT.
The energy deposited in the scintillators was smeared by considering 
the position dependence of the light collectivity and photo-electron statistics.  

Figure~\ref{fig:Cs137} shows a digitized charge (QDC) distribution 
of the 0.662-MeV $\gamma$-ray measurement, obtained by the measurement using a 
1~MBq $^{137}$Cs source at a distance of 7~cm, together with the simulation result. 
The bump structure at ${\rm QDC}\sim 240$ corresponds to the Compton edge of the 0.662~MeV $\gamma$-ray.   
The energy was smeared in the simulation with Gaussian-distributed random numbers based on the given resolution.  
The energy resolution and the energy-to-QDC conversion factor were determined  
by $\chi^2$ minimization in the area of ${\rm QDC}>150$, 
and the resultant values were  $\sigma_E/E=4.5\%/\sqrt{E [{\rm MeV}]}$ and 543 QDC/MeV, respectively. 
The simulation with the obtained detector response  
reproduced the data in the Compton edge region.   
The reduced $\chi^2$ was further improved from 5.1 to 1.9 when we applied the smearing based on 
the position dependence of the light collectivity.

\subsection{Procedure of the measurement}
To obtain the muon-flux distribution downstream of the beam dump, 
we measured the muon coincidence rate 
by changing the vertical position of the detector at 0.5 m intervals in the observing hole.
The center of the muon detector was chosen as the detector reference point in the $y$ direction for the measurement.
The position was adjusted by connecting 0.5-m square pipe joints 
to the extension square pipe on top of the detector housing.
Using this method, we can quickly and accurately set the detector to the proper position   
while keeping its front face aligned with the beam axis. 
The muon flux was measured at 11 points in the range from $y$=$-1.23$~m to $y$=+3.77~m.

The counting rate of each detector drastically varied depending on the measurement position. 
Near the beam level ($y$=+0.27~m), a single strip of the muon detector had a counting rate of 54.9~kHz, 
while near the ground level ($y$=+3.77~m), it had a rate of only 4.2~Hz, with a 0.2~MeV threshold.  
This energy threshold corresponds to one-fourth of the energy loss of minimum ionization particles (MIP).

The monitored temperature varied from $19.7^\circ$C to $24.7^\circ$C 
depending on the measurement position;  
the temperature at the bottom of the observing hole was $6^\circ$C lower than that of the ground.
Due to MPPC's characteristics, the gain can change due to the temperature variation, typically $-1.1\%/^\circ$C.
Thus, we employed a calibration constant for each measurement position in the data analysis.

We adjusted the measurement time at each position based on the counting rate.
Measurements were taken for about one minute near the beam level  
and for 34~minutes ($y$=+3.77~m) near the ground level. 
The statistical error of the coincidence rate was less than 5\% for all measurement points.

%% file: Section3.tex
\section{Data analysis}

\subsection{Energy calibration}

\begin{figure}
\begin{center}
\includegraphics[width=7.5cm, bb=0 0 960 540, trim=4.5cm 1.0cm 2.4cm 0cm, clip]{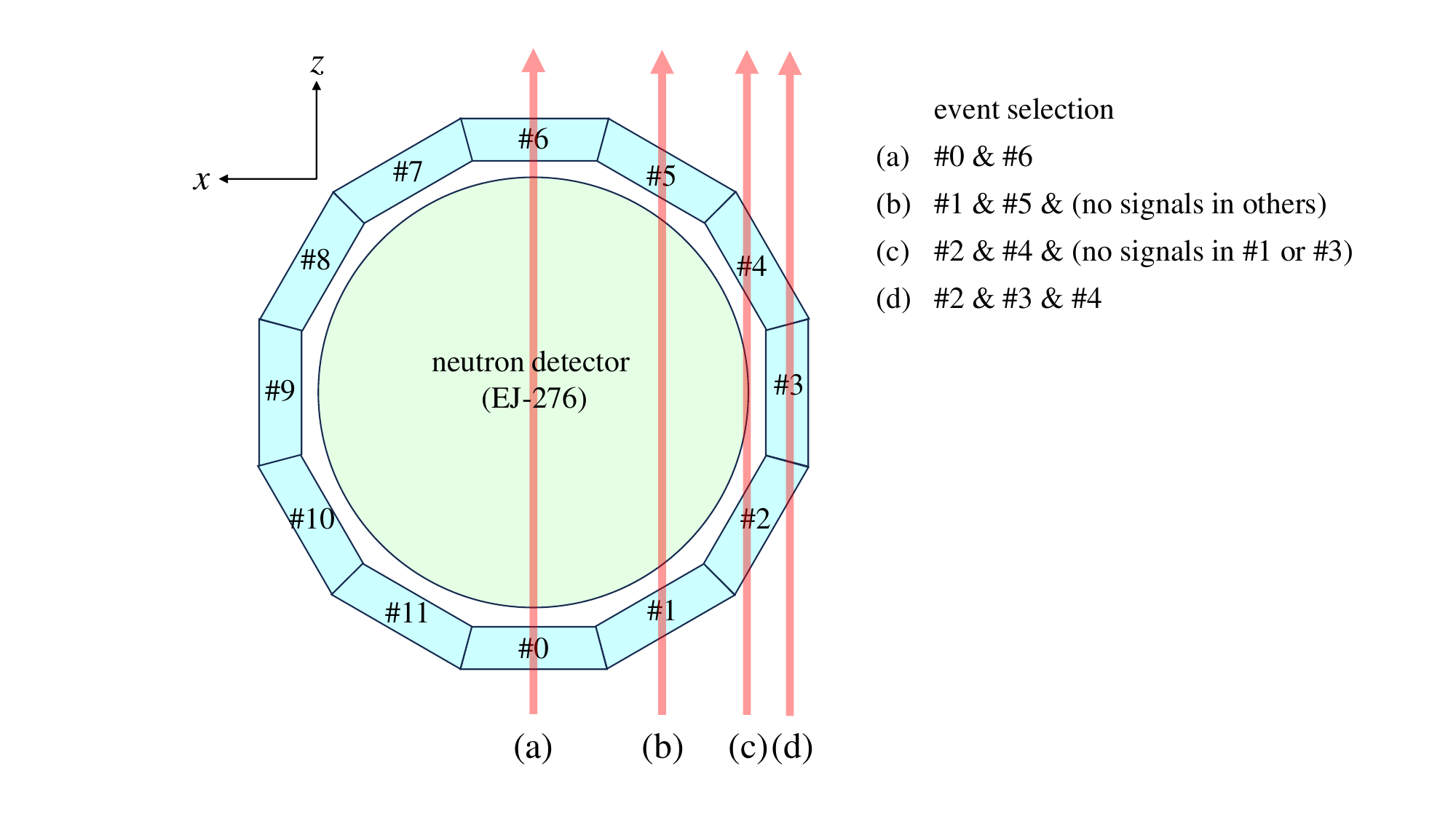}
\caption{Muon trajectories selected for energy calibration. For example, to obtain the calibration constant for strip \#2 or \#4, 
we selected the muon events with trajectory (c) to observe clear MIP peaks. Event selection for each trajectory is listed in the figure.}
\label{fig:calib_traj}
\end{center}
\end{figure}

The energy of the muon detector was calibrated by using the energy peak of MIP's that penetrate each strip.
To observe clean MIP peaks, we need to select muon events whose path length in the strip is almost constant.
As shown in Fig.~\ref{fig:calib_traj}, we selected events requiring a pattern of hit strips. 
To pick the event in trajectory~(a) in Fig.~\ref{fig:calib_traj}, the coincident hits in strips \#0 and \#6 were employed. 
The event of the trajectory~(b) was determined by the coincident hits in strips \#1 and \#5 
and the absence of signals in others.
The event of trajectory~(c) was selected by requiring coincident hits in strips \#2 and \#4 and no signals in strips \#1 and \#3.
We chose the events with triple-coincidence signal for trajectory~(d) in \#2, \#3, and \#4.

\begin{figure}
\begin{center}
\includegraphics[width=9.5cm, bb=0 0 960 540, trim=5.5cm 0cm 7.5cm 0.5cm, clip]{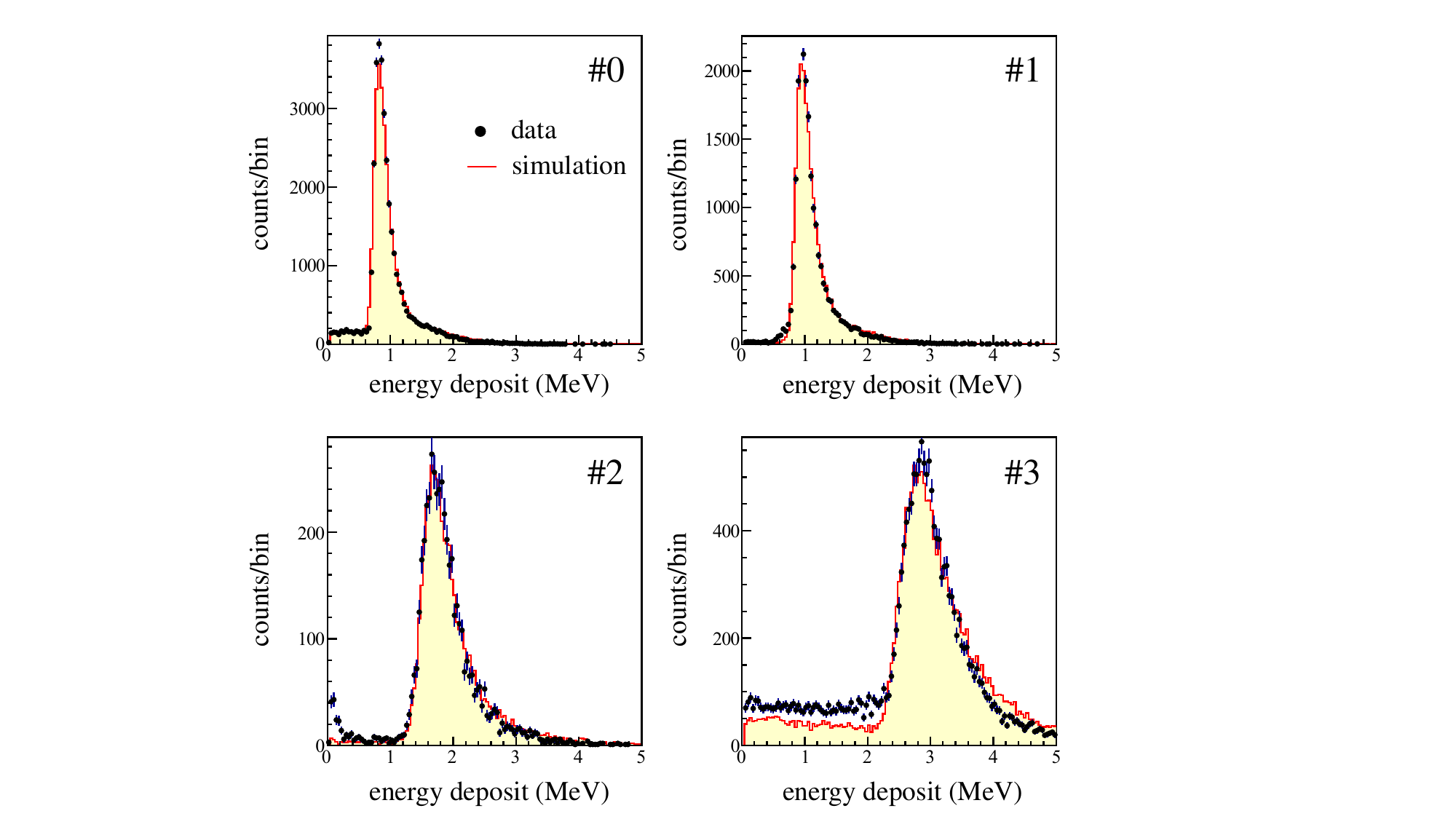}
\caption{Energy distributions for energy calibration of the muon detector. The obtained calibration constants were already applied to data in these plots. 
Histograms show the energy distributions calculated by a GEANT4 simulation with the detector response.}
\label{fig:calib}
\end{center}
\end{figure}

Figure~\ref{fig:calib} shows the energy distributions 
of strips \#0-\#3 after the event selections for the energy calibration.
Here, the QDC value of the MIP peak was already calibrated with the peak energy evaluated in the detector simulation. 
In this simulation, we defined the soil and the stainless-steel pipe of the observing hole  
to take the contribution of $\delta$-rays from surrounding substances into account and 
randomly generated muons with the energy distribution shown in Fig.\ref{fig:dump_sim}(b) 
from the average production point of $z=-16$~m.
%from the average production point of $z=-15.8$~m.
The energy distributions around the peak region in Fig.~\ref{fig:calib} were consistent between the data and the simulation results. 
A discrepancy in the high-energy tail in the distribution of strip \#3 
is caused by a dynamic-range limitation of MPPC~\cite{hpk2023}, 
which was not incorporated into the simulation response. 
The low-energy plateau in the distribution of strip \#3 was caused by low-energy $\delta$-rays 
emitted from muons of the trajectory (c); 
a slight discrepancy in this region, which is thought not to affect the energy calibration, 
would stem from the imperfect implementation of the PMT materials in the simulation.

\subsection{Selection of the muon events}

For muon event selection, we focused on forward muons and required coincident hits in strips  
\#0 and \#6 within $\pm2.5$~ns in the beam extraction period    
and no signal in any other strips during $\pm50$~ns of the coincidence timing. 
The energy threshold to define a hit was set to be 0.2~MeV. 
Contributions from the events during the beam-off period were found to be at most 0.7\% at $y$=+3.77~m  
and were subtracted from the data. 
Due to the narrow coincidence window, 
the probability of an accidental coincidence was negligibly small, less than 0.03\%.
The coincidence rate $R'$ at each measurement $y$ position is summarized in Table~\ref{tab:coin_rate}.

\begin{table*}
\caption{Summary of the measured coincidence rate $R'$, statistical error of $R'$, dead-time loss, effective detection area $S_{\rm\!eff}$, and muon flux $\phi$ for different vertical positions. 
Definitions of the variables are given in the text. The muon-flux error is the quadratic sum of the statistical error and the systematic uncertainty evaluated in Section~\ref{sec:systematic}.} \label{tab:coin_rate}
\begin{center}
\begin{tabular}{c|cccc|cc}
\hline
position  & coincidence rate & statistical & dead-time loss & effective detection & muon flux & muon-flux   \\ 
$y$ (m)              & $R'$ (s$^{-1}$)  & error  &   $1-(m_c/n_c)$    & area $S_{\rm\!eff}$ (cm$^{2}$) & $\phi$ (cm$^{-2}$s$^{-1}$) & error \\ \hline
$-1.22$ & $3.27\times10^3$ & 0.4\% &  0.4\% & 9.9  & $3.3\times10^2$  & 3.8\% \\
$-0.72$ & $9.15\times10^3$ & 0.3\% &  1.1\% & 10.0 & $9.2\times10^2$  & 3.8\%  \\
$-0.22$ & $2.34\times10^4$ & 0.2\% &  3.2\% & 10.1 & $2.4\times10^3$  & 3.8\% \\
$+0.27$ & $3.02\times10^4$ & 0.1\% &  3.7\% & 10.1 & $3.1\times10^3$  & 3.8\% \\
$+0.77$ & $1.02\times10^4$ & 0.2\% &  1.3\% & 10.1 & $1.0\times10^3$  & 3.8\% \\
$+1.27$ & $2.33\times10^3$ & 0.4\% &  0.3\% & 9.8 &  $2.4\times10^2$  & 3.8\% \\
$+1.77$ & $4.09\times10^2$ & 0.6\% &  0.1\% & 9.5 & 43                      & 3.8\% \\
$+2.27$ & 80.4                   & 0.7\% &  0.0\% & 9.3 & 8.6                     & 3.9\% \\
$+2.77$ & 14.3                   & 2.8\% &  0.0\% & 8.9 & 1.6                     & 4.7\% \\
$+3.27$ & 3.26                   & 3.3\% &  0.0\% & 8.7 & 0.38                   & 5.0\%\\
$+3.77$ & 0.696                 & 4.9\% &  0.0\% & 8.4 & 0.083                  & 6.2\% \\ \hline
\end{tabular}
\end{center}
\end{table*}

\subsection{Dead time correction}

Let the true coincidence rate be $n_{\rm c}$, the observed coincidence rate be $m_{\rm c}$, and 
the dead time of the digitizer be $\tau$. The true rate $n_{\rm c}$ can be approximately expressed 
by the following formula\footnote{
Namely, the probability of the true coincidence can be expressed as the product of three live-time probabilities:   
%the probability in the presence of noise from strip \#0 and the probability in the presence of noise from strip \#6 and 
%that in the presence of coincidence signal.}
the probabilities in the presence of noise from strip \#0 and that from strip \#6 and a coincidence signal.}~\cite{mann1956}, 
\begin{equation}
n_{\rm c}=\frac{m_{\rm c}}{[1-(m_0-m_{\rm c})\tau][1-(m_6-m_{\rm c})\tau](1-m_{\rm c}\tau)},
\end{equation}
where $m_0$ and $m_6$ denote observed single counting rates of strips \#0 and \#6, respectively.
The typical dead time of the digitizer was $\tau=0.52$~$\mu$s. 
At the position with the highest counting rate ($y$=+0.27~m), 
the loss due to the dead time, defined as  $1-(m_c/n_c)$,  
was estimated to be 3.7\% with the measured counting rates of $m_0=54.9$~kHz, $m_6=55.4$~kHz, and $m_{\rm c}=37.7$~kHz.
The dead-time loss at each measurement position is summarized in Table~\ref{tab:coin_rate}.

\subsection{Flux estimation}

Here, we discuss the muon-flux calculation. 
To obtain the flux, the true coincidence rate after the dead time correction, $R$ (=$R'\cdot n_c/m_c$), 
was divided by the apparent area ($S$) of the front (\#0) and rear (\#6) strips   
viewed from the average muon production point 
$(x,y,z)=(0,0,-16~{\rm m})$. 
%$(x,y,z)=(0,0,-15.8~{\rm m})$. 
The apparent area $S$ slightly changes depending on the measurement point,  
11.8~cm$^2$ at the beam level ($y$=0), and 9.4~cm$^2$ near the ground ($y$=+3.77~m). 
In addition, two acceptance corrections were applied based on the study with the detector simulation. 
One is a correction for acceptance loss due to multiple scattering $\varepsilon_{\rm m}$,  
which is the correction for the muons that are not on a straight trajectory and cannot be counted by the two strip layers. 
This was typically about 12\%. 
The other is the acceptance loss due to 0.2~MeV energy threshold $\varepsilon_{\rm th}$, estimated to be 0.2\%. 
We defined the effective detection area $S_{\rm\!eff}$ as 
\[
S_{\rm\!eff} = S(1-\varepsilon_{\rm m})(1-\varepsilon_{\rm th})
\]
and obtained the value $S_{\rm\!eff}$ as a whole by the simulation, 
with which the flux $\phi$ can be calculated by $\phi\ [{\rm cm^{-2}s^{-1}}]=R\ [{\rm s^{-1}}]/S_{\rm\!eff}\ [{\rm cm^2}]$\footnote{
%Specifically, we individually estimated the flux $\phi$ in the simulation by counting muons at the detector positions with a virtual sphere of 7.5~cm diameter and 
%the coincidence rate $R$ at that position with the detector response simulation. Then, we calculated the factor $S_{\rm\!eff}$ by dividing $R$ by $\phi$ at each position.
Specifically, we performed simulations with two different setups individually in order to obtain $S_{\rm\!eff}$. 
One is the simulation in which a virtual sphere of 7.5~cm diameter is placed at the detector position, and the number of muons 
entering the sphere is simply counted to determine the muon flux $\phi$ (cm$^2$s$^{-1}$). 
Another one is the simulation in which the actual detector is placed at that position and calculate number of detection by taking into account the detector's response and the 0.2-MeV threshold to determine the muon counting rate $R$ (s$^{-1}$). 
By dividing the calculated $R$ by $\phi$ at each position, we obtained $S_{\rm\!eff}$ without calculating 
$S$, $\varepsilon_{\rm m}$ and, $\varepsilon_{\rm th}$ separately. 
}. 
The effective detection area $S_{\rm\!eff}$ and obtained muon flux $\phi$ are summarized in Table.~\ref{tab:coin_rate}.

%% file: Section4.tex
\section{Systematic uncertainties}

\label{sec:systematic}

The main contribution to the systematic uncertainties in the muon-flux measurement was attributed to   
the uncertainty of the detector acceptance due to the misalignment  
of the muon-detector strips inside the aluminum protective housing.
Since the width of the strip is narrow ($\sim$16~mm) 
compared to the distance between the front and rear strips (60~mm), slight lateral deviations 
may have a sizable impact on the coincidence probability. 
This uncertainty was evaluated by utilizing the fact that the muon detector 
has 30-degree rotational symmetry around the vertical axis;  
when we compared the muon coincidence rate 
at rotation angles of 0, 30, 90, and 120 degrees  
under the identical position and beam conditions\footnote{These special measurements were 
carried out for 1 minute at each angle with a 50~kW beam at $y=+1.77$~m.}, 
the variation of the resultant fluxes was 3.8\%, which we quoted as the main systematic uncertainty. 
Other contributions, 
such as uncertainty in the detector solid angle coming from beam shift (0.35\%) and    
machining accuracy of the scintillator strips (0.26\%), 
the accidental coincidence caused by $e^\pm$ or $\gamma$ (0.2\%),  
the error coming from energy calibration and variation of scintillator thickness ($<$0.05\%),   
were negligibly small compared to the systematic uncertainty due to the misalignment.  

%% file: Section5.tex
\section{Results}

\begin{figure}[!t]
\begin{center}
\includegraphics[width=8.5cm, bb=0 0 960 540, trim=5.0cm 1.5cm 8.5cm 2.cm, clip]{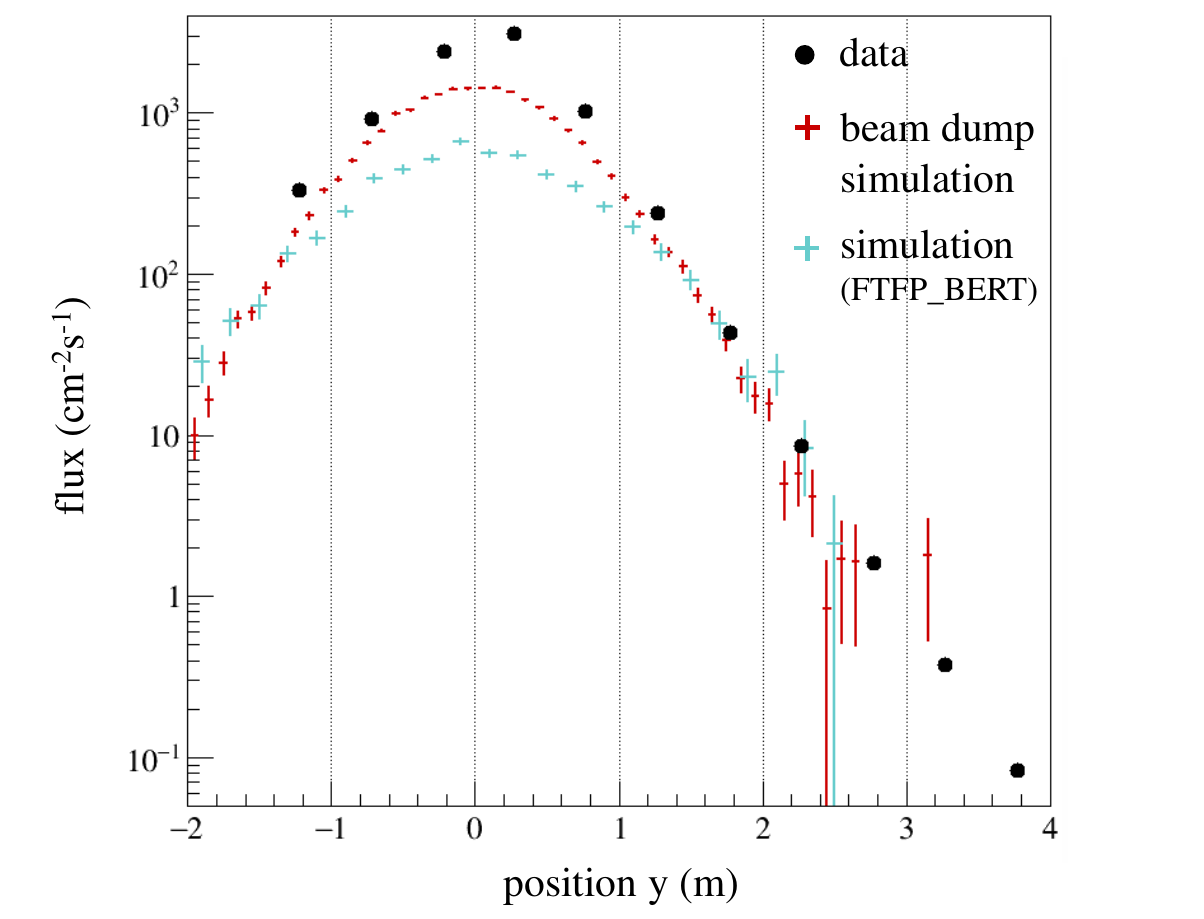}
\caption{Vertical position dependence of the muon flux at the observing hole. 
Black circles denote the measured muon fluxes with the 30~kW proton beam intensity, 
and red crosses show the muon flux calculated with the beam-dump simulation described in Section~\ref{subsec:expectation}. 
Green crosses indicate the outputs from an alternative beam-dump simulation by using a 
 different hadron physics package in GEANT4, FTFP\_BERT, instead of QGSP\_BERT, which is discussed in Section~\ref{subsec:difference}.}
\label{fig:flux}
\end{center}
\end{figure}

Figure~\ref{fig:flux} shows the vertical position dependence of 
the muon flux at the observing hole, obtained by both data and the simulation.  
As expected, the muon flux was the highest near the beam level 
and decreased rapidly in response to the distance from the beam axis.

%The simulation result shows an underestimation of up to 1/2 near the beam level.  
The simulation leads to an underestimation of a factor 2 near the beam level.  
In contrast, the calculation agrees with the data within 30\%
at positions higher than $y$=+1~m.  
For the purpose of the muon-flux estimation for the KOTO~II experiment, 
which will be placed at the horizontally off-axis, 
we can conclude that the beam dump simulation can be used with reasonable precision.

%% file: Section6.tex
\section{Discussions}

 \label{subsec:difference}

Here, we investigate why the muon flux calculated  
by the beam-dump simulation is underestimated near the beam level.

One possibility is the overestimation of the beam size of the incident protons at the beam dump.
In the beam-dump simulation  
the beam size was set to $\sigma=3.0$~cm 
based on the RGIPM measurement, as described in Section~\ref{subsec:beam}. 
Suppose the beam size was narrower than that of the measured value;  
the number of muons passing through the shield 
could increase because they may be generated further downstream 
at the conical hole of the copper core.
In addition, the RGIPM used for the measurement was not equipped 
with a permanent magnet to suppress the diffusion of electron-ion pairs~\cite{sato2012}; 
thus, the beam size may be overestimated due to its performance limitation.
Nevertheless, the number of muons would only increase approximately 1.2 times,  
even with the beam size being narrowed to 2~cm.   
In addition, the beam size evaluated with a beam optics calculation was not so narrow, 2.7~cm in $x$ and 2.8~cm in $y$ directions. 
An overestimate of the beam size, if it existed, could contribute to the underestimation of the muon flux.  
Still, the difference is not significant enough to account for the difference of factor 2.

A study has shown that inhomogeneities in shielding materials 
such as concrete and soil can lead to uncertainties in muon intensity 
behind the beam dump~\cite{battaglieri2019}.
Based on the assumption that concrete density ($\rho_{\rm conc}$) can fluctuate by $\pm 10\%$, 
we conducted simulations with densities of $\rho_{\rm conc}=2.1$~g/cm$^3$ and $\rho_{\rm conc}=2.5$~g/cm$^3$.
The number of muons that reached the observing hole varied 
uniformly regardless of the distance from the beam axis
by +24\% and -16\%, respectively, comparing to the case of nominal density of 2.3~g/cm.
Therefore, the decrease of factor 2 near the beam axis 
cannot be explained solely by the non-uniformity of the concrete.
Regarding soil uniformity, it is worth considering density change since 
the groundwater level near the observing hole was just around the beam level ($y=-0.74$~m).
However, the impact was thought to be small since the soil thickness was only 1.95~m.  
If we assume that the pores in the soil are saturated with water and the soil density ($\rho_{\rm soil}$) became to 1.85~g/cm$^3$,   
the muon reduction rate was estimated to be $-4.1\%$ compared to value calculated with the original density of $\rho_{\rm soil}=1.6$~g/cm$^3$.
This effect also cannot explain the decrease of factor 2, 
but it would be one of the reasons for the asymmetry across the beam level observed in the experimental data.

Another possibility is the ambiguity of the hadronic-physics model 
used in the GEANT4 simulation.  
The angular distribution and yield of charged pions produced 
by the interaction of the primary proton beam and copper in the beam dump 
largely affect the muon-flux distribution.  
The hadron physics package of QGSP\_BERT was used 
in the beam-dump simulation, which treats hadron interactions 
with the Quark Gluon String (QGS) model for protons above 12~GeV, 
Fritof (FTF) parton model in the region of 3-12~GeV and with Bertini Cascade model below 6~GeV~\cite{allison2016}. 
When we changed the model to FTFP\_BERT, 
which uses the FTF model for all regions above 3~GeV,   
the position dependence of the muon flux was slightly expanded and the total muon yield 
at the observing hole decreased by a factor of 2/3, as shown in Fig.~\ref{fig:flux}. 
Although we could not find the hadron package that can reproduce our measurement, 
the ambiguity of the hadronic-physics model indeed affects  
the difference in the muon flux between the data and the beam-dump simulation. 

It is crucial to verify which physics model accurately reproduces the experimental data 
for radiation protection at high-energy proton accelerator facilities.  
In the case of an intense beam facility like J-PARC, 
activation of soil due to muon capture reactions 
may affect dose assessment in terms of the dispersion of radioactive materials 
into the environment through groundwater\footnote{
For this reason, a strict dose limit for muons in soil has been set at J-PARC as a design standard for soil activation.}.  
From this point of view, it is important to conduct systematic research to reproduce our data  
using not only GEANT4 but also computational codes such as FLUKA~\cite{Ferrari2005}, MARS~\cite{mokhov2017} and PHITS~\cite{sato2024}, 
which are in standard use for shielding design in accelerator facilities. 
This research will be conducted elsewhere as the next step.

%Although this is not the purpose of this article, 
In addition, 
it should be noted that measurements of particle beam behind a beam dump at high-intensity accelerator facilities 
attract much attention as the experiments for dark matter searches \cite{alviggi2022, battaglieri2024}. 
We anticipate that our measurement will provide valuable insights in designing future beam-dump experiments.